\documentclass[prl,floatfix,twocolumn,showpacs,amsmath,amssymb]{revtex4}
\usepackage{graphicx}
\usepackage{dcolumn}
\usepackage{bm}
\usepackage[usenames]{color}
\newcommand{\dagga}{{\phantom{\dagger}}}
\begin{document}

\title{Characterization of the Bose-glass phase in low-dimensional lattices}
\author{Juan Carrasquilla,$^{1}$ Federico Becca,$^{1,2}$ 
Andrea Trombettoni,$^{1,3}$ and Michele Fabrizio$^{1,2,4}$}
\affiliation{
$^{1}$ International School for Advanced Studies (SISSA), Via Beirut 2, I-34151, Trieste, Italy \\
$^{2}$ Democritos Simulation Center CNR-IOM Istituto Officina dei Materiali, Trieste, Italy \\
$^{3}$ INFN, Sezione di Trieste \\
$^{4}$ International Centre for Theoretical Physics (ICTP), P.O. Box 586, I-34014 Trieste, Italy}
\date{\today}

\begin{abstract}
We study by numerical simulation a disordered Bose-Hubbard model in 
low-dimensional lattices. We show that a proper characterization of the phase 
diagram on finite disordered clusters requires the knowledge of probability 
distributions of physical quantities rather than their averages. This holds 
in particular for determining the stability region of the Bose-glass phase, 
the compressible but not superfluid phase that exists whenever disorder is 
present. This result suggests that a similar statistical analysis should be 
performed also to interpret experiments on cold gases trapped in disordered 
lattices, limited as they are to finite sizes.
\end{abstract}


\maketitle

{\it Introduction --}
The impressive progresses in experiments with ultra-cold gases trapped in 
optical lattices have revived interest in old yet fundamental issues of 
many-body physics.~\cite{blochrev} In fact, these systems give the unique 
opportunity to experimentally realize simple many-body models, like the Bose 
or Fermi Hubbard models, which are believed to capture the essential physics 
underneath important phenomena, like for example superfluidity or the Mott 
metal-insulator transition. 

One of the first successes of these experiments has been the observation of a
superfluid to Mott insulator transition in bosonic atoms trapped in optical 
lattices upon varying the relative strengths of interaction and inter-well 
tunneling.~\cite{bloch} The possibility of introducing and tuning disorder, 
through speckles or additional incommensurate lattices, also led to the 
observation of Anderson localization for weakly interacting Bose 
gases.~\cite{aspect,inguscio} These important achievements progressively opened
the way towards the challenging issue of realizing and studying a Bose-Hubbard
model in the presence of disorder. Preliminary attempts to measure the 
excitation spectrum of interacting bosons in a disordered 
lattice,~\cite{fallani} have been performed by using Bragg 
spectroscopy.~\cite{stoferle}

The phase diagram of a disordered Bose-Hubbard model is supposed to include 
three different phases.~\cite{giamarchi,fisher} When the interaction is strong
and the number of bosons is a multiple of the number of sites, the model 
should describe a Mott insulator, with bosons localized in the potential wells
of the optical lattice. This phase is not superfluid nor compressible. 
When both interaction and disorder are weak, a superfluid and compressible
phase must exist. These two phases are also typical of clean systems.
In the presence of disorder a third phase arises: the so-called 
Bose glass, which is compressible but not superfluid.~\cite{fisher} Indeed,
when disorder is very strong, bosons localize in the deepest potential wells,
which are randomly distributed. The coherent tunneling of a boson between 
these wells is suppressed just as in the usual Anderson localization, hence 
the absence of superfluidity, in spite of the fact that displacing a boson from
one well to another one may cost no energy, hence a finite compressibility. 
Based on the same single-particle description used for explaining Anderson 
localization, it was argued that disorder prevents a direct superfluid to Mott
insulator transition,~\cite{fisher} a speculation that has been subject to  
several theoretical 
studies.~\cite{monien,scalettar,krauth,krishna,lee,prokofiev,pollet1,pollet2}  

A simple way to justify the validity of the single-particle arguments is to 
imagine that the few carriers, which are released upon doping a Mott insulator,
effectively behave as bosons at low density. In this case the single-particle
Anderson localization scenario is likely to be applicable since the few 
interacting bosons occupy strongly localized states in the Lifshitz tails. 
The implicit assumption is that the Mott-Hubbard side bands survive in the 
presence of disorder and develop Lifshitz's tails that fill the Mott-Hubbard 
gap. This scenario is quite appealing hence worth to be investigated 
theoretically. However, a direct comparison of theory with experiments has to 
face the problem that experiments on cold gases are unavoidably limited to 
finite  systems with hundreds of sites and finite number of disorder 
realizations. Therefore, objects like Lifshitz's tails, which arise from rare 
disorder configurations, might not be easily accessible. This fact demands 
an effort to identify salient features of the Bose glass that may distinguish
the latter from a superfluid or a Mott insulator already on finite systems.

This is actually the scope of this Letter. Specifically, we are going to show
that the statistical distribution of the energy gaps extracted by a numerical
simulation of finite size systems is a significant property that can 
discriminate among different phases. The numerical simulation have been carried
out for a single chain, a two- and three-leg ladder system and finally for a
genuine two-dimensional lattice. The ladder systems are of interest because
they can be experimentally realized, not only in optical lattices but also in 
magnetic materials. Indeed, very recent neutron scattering data reported the 
evidence of the spin-analogous of a Bose-glass phase in a spin-ladder compound
in which disorder was induced by random chemical substitution.~\cite{regnault} 
Finally, we shall also discuss how the probability distribution of the energy
gaps could be experimentally accessed. 

\begin{figure}
\includegraphics[width=\columnwidth]{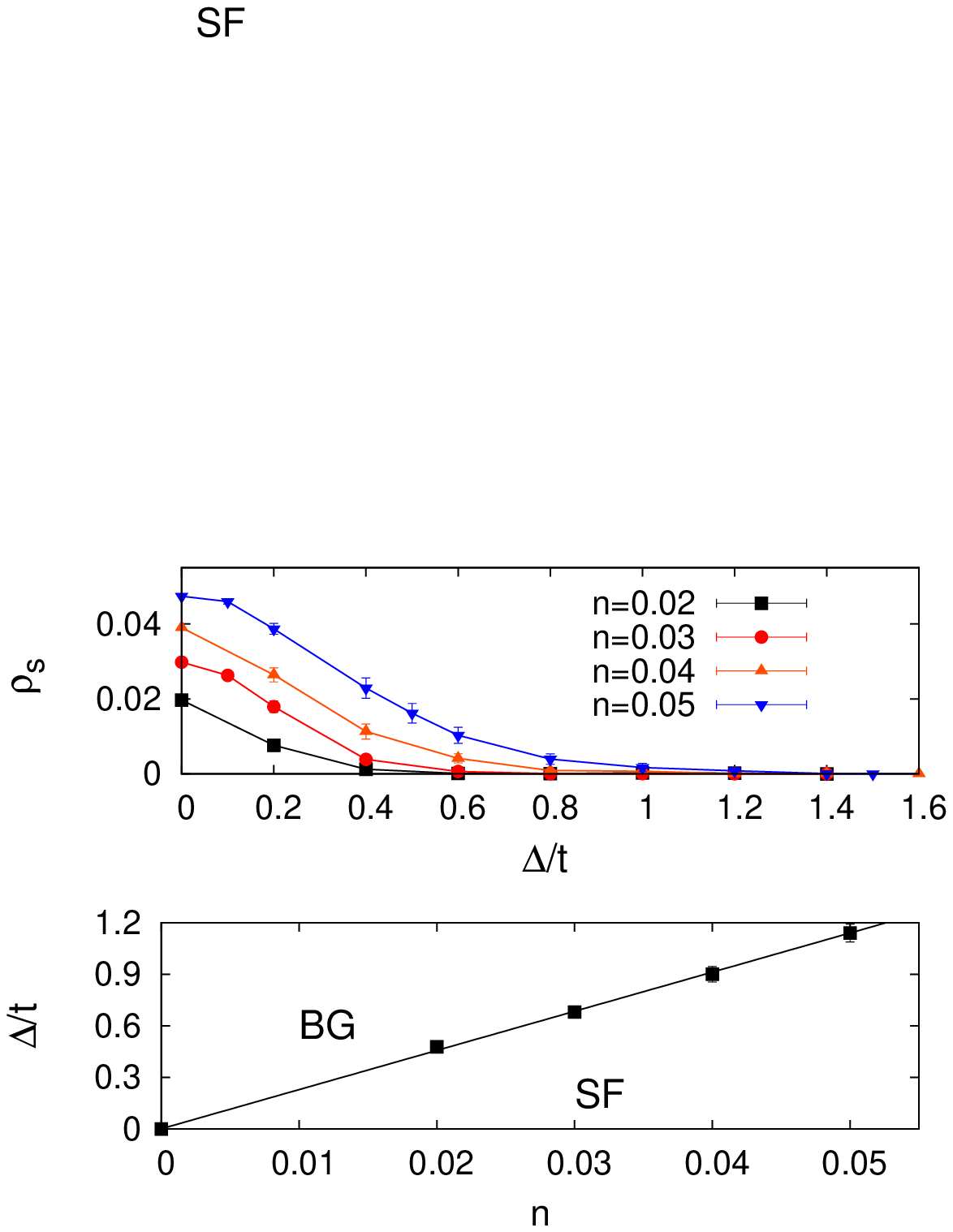}
\caption{\label{fig:hcb}
(Color on-line) Upper panel: superfluid stiffness $\rho_s$ as a function of
the disorder strength $\Delta/t$ for different densities of hard-core bosons.
Lower panel: low-density phase diagram of the hard-core bosonic model.
Calculations have been done on a $2 \times 50$ ladder system.}
\end{figure}

{\it Model --}
The simplest Hamiltonian that contains the basic ingredients of strong
correlations and disorder is
\begin{equation}\label{hambose}
{\cal H}=-\frac{t}{2} \sum_{\langle i,j \rangle} b^\dagger_i b^\dagga_j + h.c.
+ \sum_i \left( \frac{U}{2} n_i (n_i-1) + \epsilon_i n_i \right ),
\end{equation}
where $\langle \dots \rangle$ indicates nearest-neighbor sites, $b^\dagger_i$
($b^\dagga_i$) creates (destroys) a boson on site $i$, and $n_i=b^\dagger_i b_i$
is the local density operator. The on-site interaction is parametrized by $U$,
whereas the local disordered potential is described by random variables
$\epsilon_i$ that are uniformly distributed in $[-\Delta,\Delta]$.
Here, we consider bosons on a one-dimensional (1D) chain, $N$-leg 
ladders, and a two-dimensional (2D) square lattice, and study model of 
Eq.~(\ref{hambose}) by Green's function Monte Carlo with a fixed number $M$ 
of bosons on $L$ sites, $n=M/L$ being the average density. In realistic 
experimental setups, a two-leg ladder can be realized through a double well 
potential along a direction (say, $x$),~\cite{albiez} a potential creating a 
cigar geometry in the $z$-axis, and finally a periodic potential along 
$z$.~\cite{notet}

{\it Results --}
Before considering the case of finite $U/t$, let us briefly discuss the limit
of hard-core bosons (i.e., $U=\infty$) at low densities. Fig.~\ref{fig:hcb} 
shows the low-density phase diagram on a two-leg ladder. We find that for any 
finite disorder $\Delta$, the low-density phase is a Bose glass that turns 
superfluid above a critical density. In other words, the trivial Mott insulator
with zero (or one) bosons per site is indeed separated from the superfluid
phase by a Bose glass. We emphasize that the existence of a superfluid phase
for hard-core bosons in a two-leg ladder is {\it per se} remarkable. Indeed, 
in a single chain with nearest-neighbor hopping, hard-core bosons are 
equivalent to spinless fermions, which Anderson localize for any density and 
in any dimension $D\leq 2$. Consequently, hard-core bosons on a single chain 
are never superfluid. Already in a two-leg ladder, hard-core bosons start to 
behave differently from spinless fermions. Indeed, while the latter ones remain 
always localized, the former ones show a superfluid phase. We just mention that 
the same occurs also on a single chain with longer-range hopping. 

\begin{figure}
\includegraphics[width=\columnwidth]{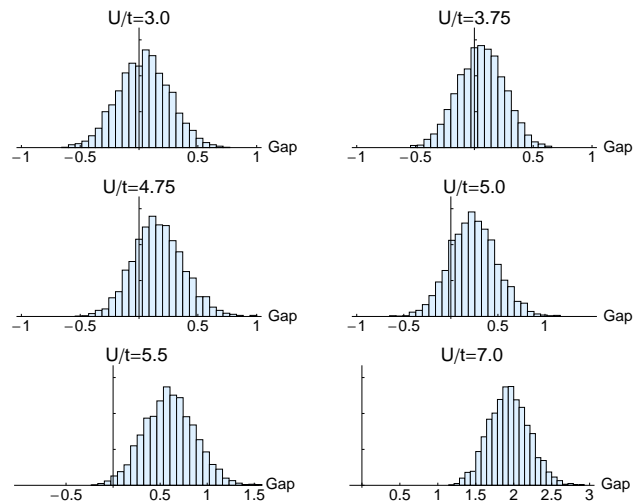}
\caption{\label{fig:histogram1d}
Distribution $P(E_g)$ of the gap in the 1D Bose-Hubbard model for different 
values of $U/t$ and $L=60$ sites.}
\end{figure}

We now turn to finite on-site interactions and consider the case with $n=1$. 
The Bose-Hubbard model has been extensively studied in recent years,
~\cite{monien,scalettar,krauth,krishna,lee,prokofiev,pollet1,pollet2} with 
special focus on the question whether a direct superfluid to Mott insulator 
transition does exist or not. This issue has been finally solved only recently. 
The solution is based on the observation that, if the disorder strength 
$\Delta$ is larger than half of the energy gap of the clean Mott insulator 
$E_{g}^{\rm clean}$, then the ground state must be  compressible, otherwise 
is incompressible.~\cite{pollet1,weichman} Therefore, the independent 
measurements of the superfluid stiffness $\rho_s$ at finite $\Delta$ 
and of the clean Mott gap $E_{g}^{\rm clean}$ allow a precise determination of
the phase boundaries between different phases and demonstrate unambiguously
the existence of a Bose glass in between the superfluid and Mott 
phases.~\cite{pollet1,pollet2} The above prescription is very effective in a 
numerical simulation since both $\rho_s$ with disorder and $E_{g}^{\rm clean}$
without disorder can be determined quite accurately. On the other hand, it
would be desirable to have simple instruments to establish directly the nature
of the phase of a given system in a realistic finite-size experimental setup. 
In a clean system, this program can be accomplished by measuring the gap, 
conventionally defined by $E_{g}=\mu^{+}-\mu^{-}$, where 
$\mu^{+}=E_{M+1}-E_{M}$ and $\mu^{-}=E_{M}-E_{M-1}$ ($E_{M}$ being the 
ground-state energy with $M$ particles). Experimental estimates for the gap 
have been so far obtained in ultra-cold atomic systems mainly in two ways: 
one consists in applying a gradient potential that compensates the Mott energy
gap and  allows tunneling between neighboring sites;~\cite{bloch} the other 
method exploits a sinusoidal modulation of the main lattice height for 
stimulating resonant production of excitations.~\cite{stoferle,fallani}

In disordered systems, the Mott gap can be overcome by transferring particles 
between two regions with almost flat disorder shifting the local chemical 
potential upward and downward, respectively. These regions may be far apart 
in space and represent rare fluctuations (Lifshitz's tail regions).
Therefore, it is quite likely that the conventional definition of the gap,  
$\bar{E}_g = 1/{\cal N} \sum_{\alpha=1,\dots,{\cal N}} \left (
\mu^+_\alpha-\mu^-_\alpha \right )$, where $\alpha$ denote the disorder
realizations, will miss the Lifshitz's tails for any accessible number of
disorder realizations ${\cal N}$. This fact gives rise to a finite gap, even
when the actual infinite system would be compressible.         
To circumvent such a difficulty, it is useful to imagine that a large systems
is made by several subsystems, each represented by the $L$-site cluster under
investigation, and construct the gap by using $\mu^{+}$ and $\mu^{-}$ from 
{\it different} disorder realizations. In other words, one could define an 
alternative estimate of the gap as $E_g^{{\rm min}}={\rm min}_{\alpha,\beta} 
\mid \mu^{+}_\alpha - \mu^{-}_\beta \mid$, with all the disorder 
realizations $\alpha$ and $\beta$. In the limit of very large systems where 
boundary effects become negligible, $E_g^{{\rm min}}$ must eventually 
coincide with $\bar{E}_g$. In finite systems the two estimates differ, 
nevertheless we believe that $E_g^{{\rm min}}$ is more representative since 
it can capture the phenomenon underneath the Lifshitz's tails, as we are going
to show numerically. Besides $E_g^{{\rm min}}$, one can determine the full 
gap distribution, 
$P(E_g) = \sum_{\alpha\beta}\,\delta \left (
E_g-\mu^{+}_\alpha + \mu^{-}_\beta \right )$, 
which we will show has remarkable properties. We mention that, by our 
definition, $P(E_g<0)$ could well be finite on finite systems, although it must
vanish in the thermodynamic limit where $P(E_g)$ becomes peaked at a single 
positive (or vanishing) value, i.e., the actual gap. In experiments with 
ultra-cold atoms, both $E_g^{{\rm min}}$ and $P(E_g)$ could be accessed by 
measuring {\it separately} $\mu^{+}$ and $\mu^{-}$ for different disorder 
realizations. For instance, one could measure the energy releases 
$E^{{\rm rel}}_M$ of falling atoms when the trap is turned off with the 
reference number of particles $M$ and with numbers $M \pm M^\prime$. 
For $M^\prime \ll M$, indeed
$E^{{\rm rel}}_{M+M^\prime}-E^{{\rm rel}}_{M} \simeq M^\prime \mu^{+}$ 
and 
$E^{{\rm rel}}_{M}-E^{{\rm rel}}_{M-M^\prime} \simeq M^\prime \mu^{-}$. 

\begin{figure}
\includegraphics[width=\columnwidth]{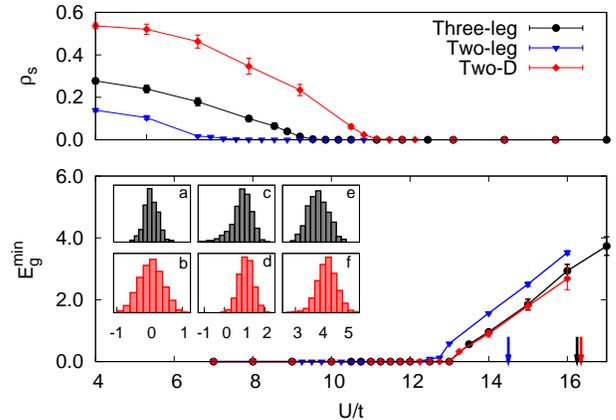}
\caption{\label{fig:phasediag}
(Color on-line) Upper panel: superfluid stiffness $\rho_s$ for different
clusters. Two-leg (with $2 \times 40$ sites) and three-leg (with $3 \times 50$)
ladders are shown; the 2D case with a $12 \times 12$ cluster is also reported
for comparison. In all cases the disorder strength is $\Delta/t=5$.
Lower panel: the same as in the upper panel for the minimum gap 
$E_g^{{\rm min}}$. Arrows indicate the opening of the charge gap according
to $\Delta=E_{g}^{\rm clean}/2$. The histograms for the gap are also reported
for the three-leg ladders (upper raw) and 2D (lower raw): $U/t=7$ (a and b),
$13$ (c and d), and $16$ (e and f).}  
\end{figure}

Let us start from the 1D case, whose zero-temperature phase diagram has been
worked out by Density-Matrix Renormalization Group 
(DMRG).~\cite{schollwock} At finite values of $\Delta$, the on-site interaction
$U$ turns the Bose glass into a superfluid, which remains stable up to
$U=U_{c1}$, where $\rho_s$ vanishes. However, the system remains gapless for 
$U_{c1}<U<U_{c2}$, indicating the presence of a Bose-glass phase. 
At $U=U_{c2}$ the system turns into an incompressible Mott insulator.
For $\Delta/t=2$, we have that $U_{c1}/t\simeq 3.7$. If we use $\bar{E}_g$ 
as estimator of the actual gap, we find that the Bose glass survives up to 
$U_{c2}/t \simeq 5$, not far from the DMRG estimate,~\cite{schollwock}
but smaller than the value predicted by the condition 
$\Delta=E_{g}^{\rm clean}/2$, which would lead to $U_{c2}/t \simeq 6.9$.
As discussed before, this discrepancy arises by the inability to catch rare 
disorder configurations, which could be overcome by analyzing the minimum gap
$E_g^{{\rm min}}$ and the full distribution probability $P(E_g)$.
Indeed, when using $E_g^{{\rm min}}$ as a detector of gapless excitations,
we obtain an estimate of $U_{c2}/t \simeq 6.2$, much closer to the value 
$U_{c2}/t \simeq 6.9$. As far as $P(E_g)$ is concerned, we note that it behaves
quite differently in the three different phase, see Fig.~\ref{fig:histogram1d}.
As long as the phase is superfluid, $P(E_g)$ is peaked at $E_g=0$.
In the Bose glass, $P(E_g)$ is instead peaked at a finite $E_g>0$, yet $P(0)$
stays finite. In the Mott insulator, $P(E_g)$ remains peaked at a positive
$E_g$ but $P(0)=0$. This suggests that $P(E_g)$ could be an efficient tool
for discriminating between the different phases.
  
\begin{figure}
\includegraphics[width=\columnwidth]{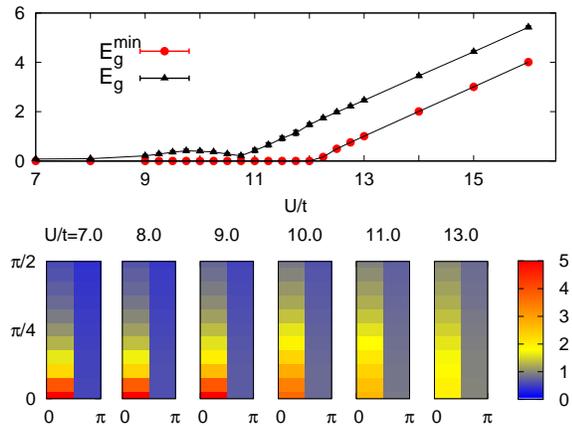}
\caption{\label{fig:momentum}
(Color on-line) Upper panel: variational results for the excitation gap for
a $2 \times 40$ ladder. Lower panels: momentum distribution $n_k$ for the same
cluster.}
\end{figure}

Let us now analyze the evolution of the phase diagram when the 2D limit is
approached by increasing the number of legs. Moving from $D=1$ to $D=2$, 
the stability region of the Bose glass is expected to shrink,~\cite{fisher} 
making its observation in experiments more and more difficult. 
In Fig.~\ref{fig:phasediag}, we report our results for two- and three-leg 
ladders, and for comparison, also the 2D limit (evaluated for a rather small 
$12 \times 12$ cluster). In this case, we take $\Delta/t=5$, in order to have
a larger Bose-glass region in between the superfluid and the Mott phases. 
In 1D, for such large disorder strength no superfluidity is found at all. 
By increasing the number of legs, we rapidly converge to the 2D results: 
this fact is particularly clear from the data on the gap. Both the results 
on the minimum gap and the ones that come from $\Delta=E_{g}^{\rm clean}/2$ 
shows that the critical $U$ for the Mott transition is almost the same for 
three legs and 2D. Also the superfluid stiffness $\rho_s$ seems to rapidly 
converge from below to the 2D limit. We also find that the behavior of 
$P(E_g)$ is qualitatively similar to what found in 1D, confirming that it can 
actually discriminate among the different phases. We mention that, should we 
use as estimator of the gap $\bar{E}_g$, we would have concluded that the 
Bose glass never exists in 2D and that a direct superfluid to Mott insulator 
transition occurs. The use of $E_g^{{\rm min}}$ instead demonstrates that 
the Bose glass does exist also in 2D and always intrudes between the superfluid
and the Mott insulator.   

We finish by showing variational results for the momentum distribution
$n_k=\langle b^\dagger_k b^\dagga_k\rangle$, obtained by the technique
outlined in Ref.~\cite{capello}. We just recall that this variational 
approach is based upon a Jastrow wave function and is able to describe equally
well superfluid, Bose-glass, and Mott insulating states. 
In Fig.~\ref{fig:momentum}, we show the results for a $2 \times 40$ ladder 
and different values of $U/t$ (we also report the results for the variational
gap). Since, this is an almost 1D system, no condensation fraction is found 
(i.e., $n_0/L \to 0$ in the thermodynamic limit). However, the superfluid 
phase is characterized by quasi-long-range order with a cusp in $n_k$ and a 
logarithmic divergent $n_0$. On the other hand, both the Bose-glass and the 
Mott phases have a smooth momentum distribution, with $n_0 \to {\rm const}$ 
in the thermodynamic limit.

{\it Conclusions --}
We have presented a detailed study of the ground-state properties of the
disordered Bose-Hubbard model in low-dimensional lattices, relevant for 
on-going experiments with cold atomic gases trapped in optical lattices. 
We have determined the distribution probability of the gap on finite sizes 
and shown that it contains useful information. In particular, we have found 
that the Bose-glass is characterized by a broad distribution of the gap that 
is peaked at finite energy but extends down to zero, a shape remarkably 
reminiscent of preformed Hubbard sidebands with the Mott gap completely 
filled by Lifshitz's tails. The Mott transition occurs when these tails 
terminate at finite energy. On the contrary, the gap distribution in the 
superfluid phase turns out to be strongly peaked at zero energy. These results
suggest a simple and efficient way to discriminate between different phases
in experiments, which, being performed on finite systems, suffer from the 
same size limitations as our simulations.   

We have also investigated the disordered Bose-Hubbard model on $N$-leg ladder
systems, emphasizing that these geometries could be quite useful to study
the evolution from one to two spatial dimensions. Experiments with both cold 
atomic gases and magnetic systems are becoming now possible on ladders and 
our calculations represent an important benchmark in this direction. 

We thank C. Castellani, L. Fallani, and C. Fort for useful discussions.
Calculations have been performed on the cluster Matrix of CASPUR, thanks to 
Standard HPC Grant 2009.


\begin{thebibliography}{99}
\bibitem{notet} Similar values of inter- and intra-chain hopping parameter
   may be achieved by an appropriate tuning of the lattice spacings and 
   height of barriers.
\end{thebibliography}
\end{document}